\documentclass{an}
\usepackage{graphicx}
\usepackage{times}
\usepackage{natbib}
\begin{document}

\bibliographystyle{aa}

\Pagespan{1009}{1012}
\Yearpublication{2007}%
\Yearsubmission{2007}%
\Month{10}%
\Volume{328}%
\Issue{10}%
\DOI{10.1002/asna.200710837}%

\title{Meridional flow profile measurements with SOHO/MDI}

\author{U. Mitra-Kraev\thanks{Corresponding author:
  \email{u.mitrakraev@sheffield.ac.uk}\newline}
\and  M. J. Thompson
}
\titlerunning{Meridional flow profile measurements with SOHO/MDI}
\authorrunning{U. Mitra-Kraev \& M. J. Thompson}
\institute{University of Sheffield, Department of Applied Mathematics,
  Hicks Building, Sheffield, S3 7RH, UK} 

\received{2007 Sep 9}
\accepted{2007 Oct 30}
\publonline{2007 Dec 15}


\abstract{We present meridional flow measurements of the Sun using a novel
helioseismic approach for analyzing SOHO/MDI data in order to push the
current limits in radial depth. 
Analyzing three consecutive months of data during solar minimum, we
find that the meridional flow is as expected poleward in the upper
convection zone, turns equatorward at a depth of
around 40\,Mm ($\sim0.95\,R_\odot$), and possibly changes
direction again in the lower convection zone. This may indicate two
meridional circulation cells in each hemisphere, one beneath the
other. 
}

\maketitle

\section{Introduction}
The first observations of the Sun's meridional flow were made independently
by \citet{Duvall1979}, \citet{Howard1979}, and \citet{Beckers1979}, who
found a poleward drift of $\sim$20\,m/s at the solar surface.
By mass conservation, this meridional flow has to change direction somewhere in the solar interior, which may happen in the lower regions of the convection zone. 

The meridional circulation plays an important role in many solar models. 
In mean-field models of angular momentum transport, the meridional circulation is responsible for how differential rotation is established and maintained \citep{Ruediger1989,Thompson_etal2003,Miesch2005}. 
Anisotropic Reynolds stresses, arising from random turbulent fluctuations of the plasma motions, can induce a meridional circulation, which in turn drives the differential rotation. 
In numerical simulations, \citet{BrunToomre2002} find that the differential rotation profile is established by turbulent convection, and that the Reynolds stresses play a crucial role in transporting meridional momentum toward the equator.
The meridional circulation is also an important ingredient in some dynamo models \citep{Charbonneau2005}.
In flux-transport models of the solar dynamo (Babcock-Leighton models) the meridional flow is needed to transport the surface poloidal magnetic field to the bottom of the convection zone, where it is then converted into a toroidal field by rotational shear \citep{DikpatiCharbonneau1999}. 

Using helioseismic techniques, the meridional flow is usually observed in the upper layers of the solar convection zone.
Quite good spatial resolution is achieved in this regime, in particular with ring-diagram analysis \citep[e.g.,][]{GH_etal1999}.
Deeper down in the convection zone the horizontal resolution is poorer
and to probe there the viewing angle needs to be large.
No return flow has conclusively been observed to date.
Using time-distance helioseismology, \citet{Giles1999}
measured the radial profile of the meridional circulation down to the bottom
of the convection zone and, with the added con\-straint that the
meridional circulation has to be contained within the convection zone,
inferred a return flow at around 0.8 solar radii.
Applying Hankel analysis, \citet{BraunFan1998} saw a tentative return
flow at around 40 Mm depth (but also consistent with zero within the
error margin), whereas \citet{DuvallKos2001} observed no return flow.

\section{Data and analysis}
We analyze velocity data from the Michelson-Doppler Imager
\citep[MDI,][]{Scherrer_etal1995} on board the Solar and Heliospheric
Observatory \citep[SOHO,][]{DomingoFleckPoland1995}. 
The observations were taken
during three consecutive Carrington Rotations: CR 1922 (1997
April 24 -- May 21), CR 1923 (1997 May 21 -- June 17), and CR 1924
(1997 June 17 -- July 13). During this time, the Sun was at the
minimum of its activity cycle. We use tracked data along the Center
Meridian\footnote{http://soi.stanford.edu/sssc/progs/mdi/track.html},
obtained for the SOI Dynamics Campaign using high-resolution,
full-disk Dopplergrams. The cadence is 60\,s.

\subsection{Dispersion relation along the center meridian}

\begin{figure*}[ht]
\centering
\includegraphics[width=150mm]{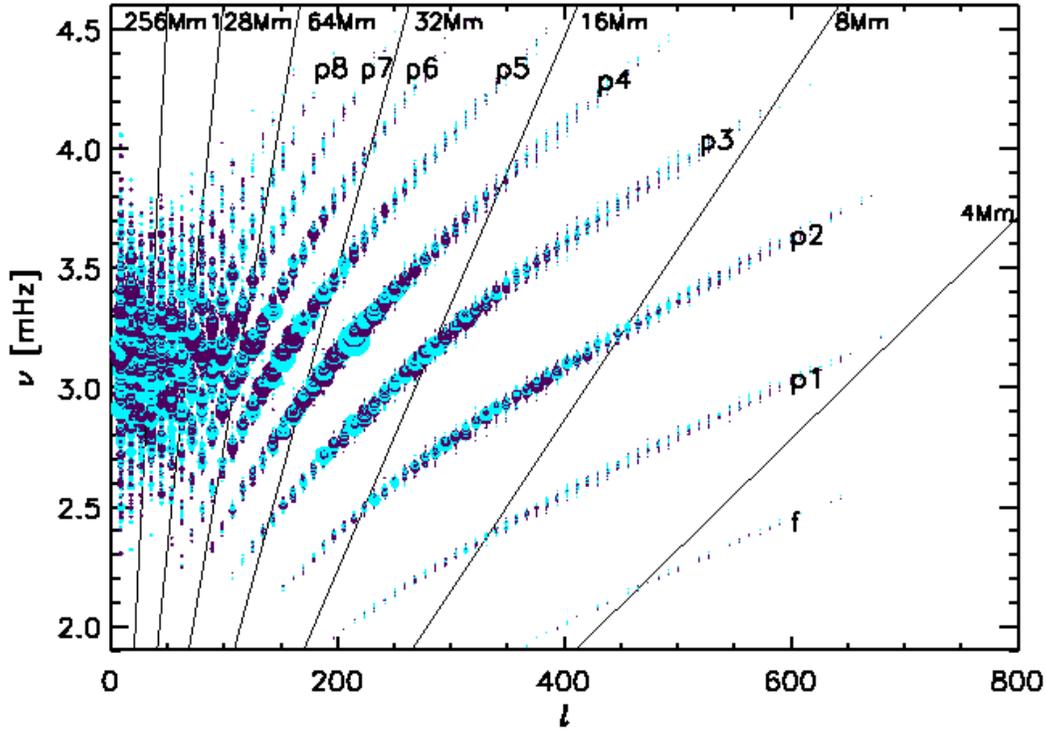}
\caption{The difference between the negative and the
  positive power spectrum, $P_{\rm neg}-P_{\rm pos}$. The frequency
  shift is visible along the f- and p-mode ridges. Dark dots are for
  $P_{\rm neg}> P_{\rm pos}$, while lighter dots are for $P_{\rm neg}<
  P_{\rm pos}$. Northward flow is present if the dark dots are shifted
  to higher frequencies and lighter dots to lower ones, while for
  Southward flow lighter dots are at higher frequencies than dark
  dots. The size of the dots is proportional to the absolute
  power. The black lines indicate the lower ray-turning points. } 
\label{fig1}
\end{figure*}

As our aim is to measure the meridional flow, which is in
North-South direction, we restrict ourselves to analyzing
spatially 1D data in the same direction.   
The data are co-added along latitude, resulting in a 2D array with one
time and one spatial coordinate \citep[see][]{UMKThompsonWoodard2006}. 
The spatial coordinate $\theta$
(latitude) is equally spaced in angle (spacing ${\rm d}\theta=0.125^\circ$). 
As the meridional flow is opposite in each hemisphere and
vanishes around the equator, the range $+20^\circ$ to $+60^\circ$ is
used for the Northern, and $-60^\circ$ to $-20^\circ$ for the Southern
hemisphere.  
For each hemisphere we obtain a power spectrum
\[
P(\nu,l) = \left| \widehat{f}(\nu,l) \right|^2,
\]
where
\[
\widehat{f}(\nu,k_\theta) = \sum f(t,\theta) \ {\rm e}^{-i(2\pi\nu t +
  k_\theta \theta)}
\]
is the Fourier transform of the data $f(t,\theta)$.
The power spectrum has symmetries such that $P(|\nu|>0,|l|>0) =
P(|\nu|<0, |l|<0)\equiv P_{\rm pos}$, which we define as the `positive', and
$P(|\nu|>0,|l|<0) = P(|\nu|<0,|l|>0)\equiv P_{\rm neg}$, which we
define as the `negative'
power spectrum. The difference between the positive and the negative
power spectrum can be attributed to flows.  

The approach of a 1D Fourier transform along the center
meridian is comparable to a spherical harmonic treatment with only
considering $m=0$ modes, which was used by \citet{KriegerRothLuehe2007} for
meridional flow measurements. As
$k_\theta$ corresponds to the spherical harmonic degree $l$, we use
the notation
$k_\theta\equiv l$ from now on. 
With the observing length $T_{\rm obs}\approx27\,{\rm days}$ and the
latitudinal range $\theta_{\rm range}=40^\circ$, the resolutions are
given by ${\rm d}\nu=1/T_{\rm obs}\approx 0.4\,\mu{\rm Hz}$ and ${\rm
  d}l = 2\pi/\theta_{\rm range}\approx 9$. 

\subsection{The effect of flows on the p-mode ridges}

For slow flows, which allow a linear approximation, the frequency
splitting is given by $\Delta\omega = k_h\cdot U^\prime$ (an
expression which is also used in ring-diagram analysis) where $k_h$
is the horizontal wave number and the velocity $U^\prime$ is the
radially-dependent horizontal flow $U(r)$ convolved with a function
describing the medium the pressure wave passes through. For measured
$\Delta\omega$ as a function of $k_h$, $U(r)$ can be retrieved by an
inversion.

\begin{figure*}[t]
\centering
\includegraphics[width=150mm]{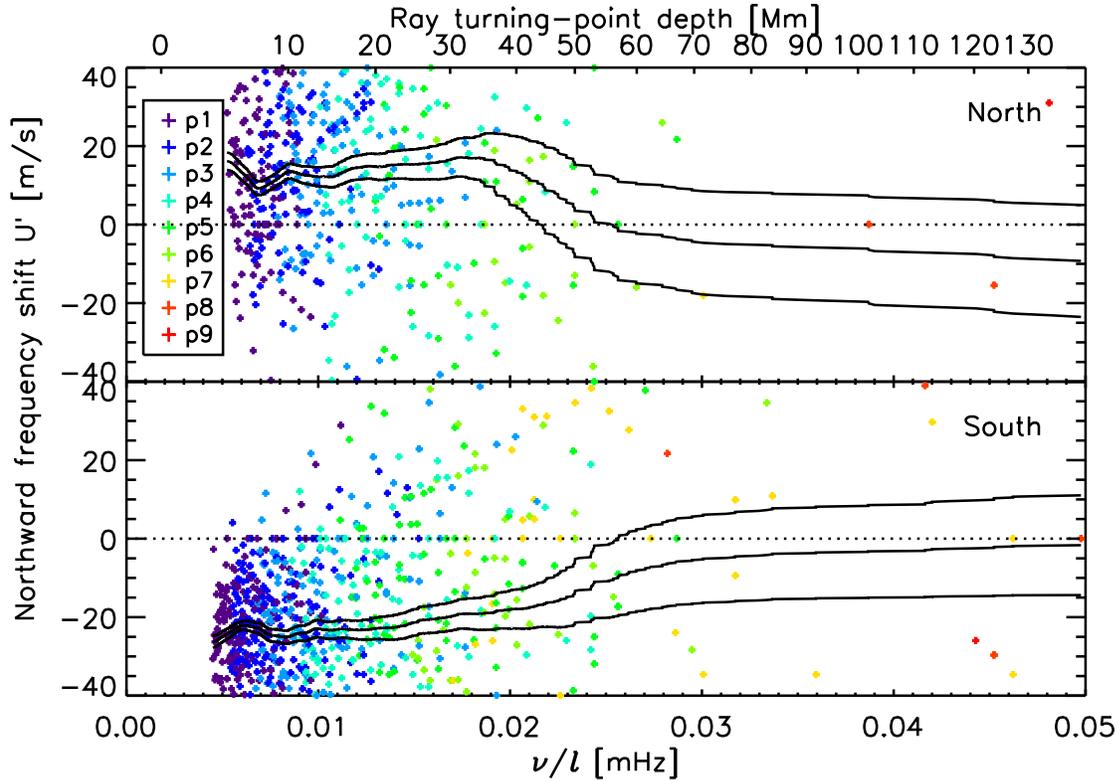}
\caption{The observed frequency shifts in the Northern (top) and
  Southern (bottom) hemispheres.} 
\label{fig2}
\end{figure*}
In our case $k_h$, $U^\prime$ and $U(r)$
are 1D variables, describing the flow perpendicular to
the solar radius in the North-South direction. 
With
$\Delta\omega = 2\pi\Delta\nu$ and $k_h=l/R_\odot$ we obtain for the
frequency shift in velocity units
\begin{equation}\label{eq:Uprime}
U^\prime(\nu/l) = \Delta\nu 2 \pi R_\odot /l,
\end{equation}
where $R_\odot=6.96\times 10^8\,{\rm m}$ is the solar radius. The
frequency shift $\Delta\nu$ is the shift by which, for a given $l$
value, a p-mode ridge is displaced from its unperturbed position due
to a flow. A Northward flow moves p-mode ridges along $l$, such that
the frequency of p-mode ridges in the positive power spectrum is
reduced by $\Delta\nu$, while the frequency of p-mode ridges in the
negative power spectrum is enhanced by $\Delta\nu$. 

Figure~\ref{fig1} shows the difference between the negative and the
positive power spectrum, $P_{\rm neg}-P_{\rm pos}$, for the Southern
hemisphere of CR 1922--1924. The size of the dots is linear with
absolute spectral power. 
For better readability, all power smaller than a
lower cutoff, which was chosen to decay exponentially along $l$ and
has a Lorentzian shape as a 
function of $\nu$ around 3.3\,mHz, was set to zero. 
The f- and p-mode ridges are clearly visible due to
the frequency shift. The f mode as well as p1--p8 modes are labeled.
 Dark dots are for $P_{\rm neg}-P_{\rm
  pos}>0$, while lighter dots represent $P_{\rm neg}-P_{\rm
  pos}<0$. This means a lighter dot being at a slightly larger
frequency than its darker counterpart indicates a Southward flow,
whereas a dark dot sitting at a larger frequency indicates a Northward
flow. The straight black lines indicate the lower ray-turning points.

\subsection{Measuring the frequency shift}

In order to obtain the frequency shift $\Delta\nu$, we cross-correlate
$P_{\rm neg}$ and $P_{\rm pos}$ for each $l$ over a $\nu$-range
covering each p-mode ridge. The criterion for accepting a measurement is a
correlation coefficient of 0.35 or more.

Figure~\ref{fig2} shows the measured frequency shift $U^\prime(\nu/l)$
for the Northern (top panel) and Southern (bottom panel) hemispheres obtained for
each power spectrum of the three 
Carrington rotations. The spread ($-200$ to $+400$\,m/s) of the individual
frequency shifts, represented by crosses and color coded with the
corresponding p-mode, exceeds the plotting range. The middle black curve
was obtained by smoothing the individual measurements twice with a bin
size of 101 points. 
For the first smoothing, the error is given by the standard deviation
of the points. The error margins in Figure~\ref{fig2} (upper and lower
lines) are then given by error propagation in the second smoothing,
assuming that the initial errors given by the standard deviation of the points are independent. However, as the initial errors are clearly not independent, this treatment undoubtedly underestimates the errors somewhat.

\subsection{Asymptotic inversion} \label{Sect:Inversion}

\begin{figure*}
\centering
\includegraphics[width=150mm]{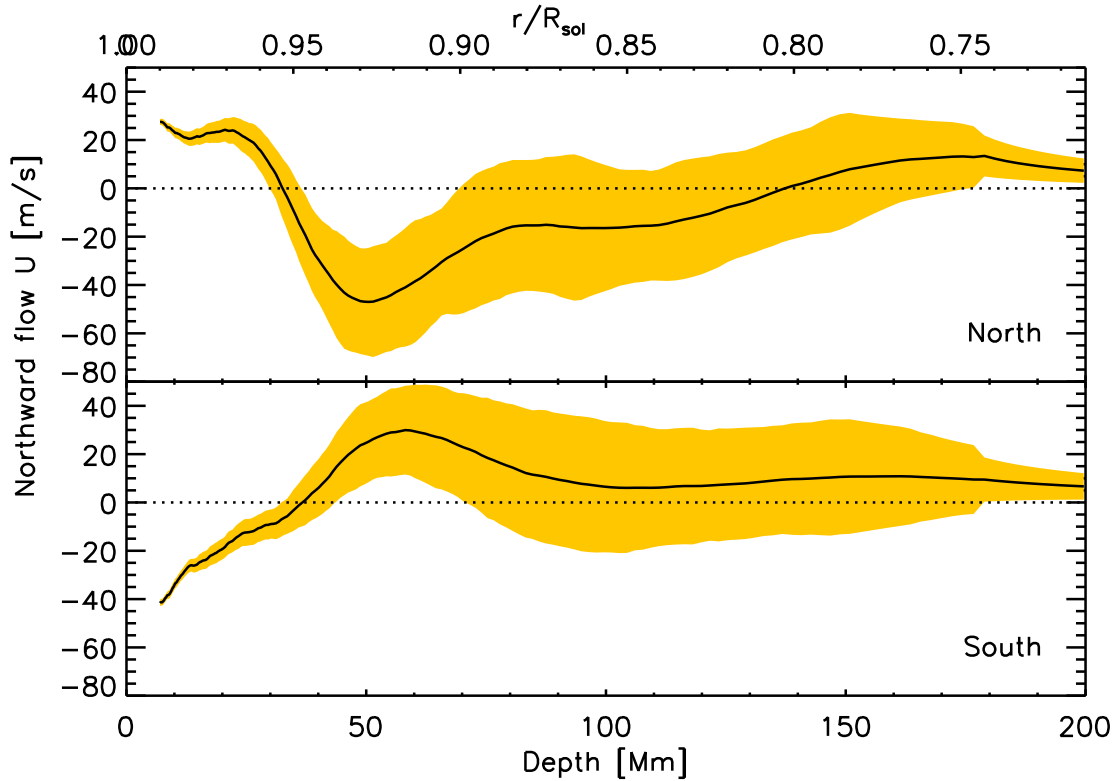}
\caption{The meridional flow profile for the Northern and
  Southern hemisphere obtained by asymptotic inversion. }
\label{fig3}
\end{figure*}
We invert the smoothed curve $U^\prime$ using an asymptotic inversion
\citep[see, e.g.,][]{JCDSchouThompson1990}. The inversion is given by 
\[
\hspace{-.5cm} U(r) = \frac{-2a(r)}{\pi} \frac{{\rm d}}{{\rm
    d\,ln}\,r} \int_{a(R_\odot)}^{a(r)}\ \sqrt{\frac{1}{a^2(r)-w^2}} \
\ U^\prime(w)\ dw 
\]
with $w=2\pi\nu/l$, $r$ the solar radius, $a(r)=c(r)/r$, and $c(r)$ the
sound-speed profile given by Model~S of \citet{JCD_etal1996}. 
The integral ranges from the
solar surface to the lower turning point of the ray path.  
In order to solve the integral, which has an integrable singularity at
the upper boundary, the numerical scheme described in Appendix II of
\citet{JCDGoughThompson1989} was applied. 

Figure~\ref{fig3} displays the obtained meridional flow profile from
the asymptotic inversion. Displayed is the Northward flow for both the
Northern and Southern he\-mispheres. The error region
(shaded area) was obtained by Monte-Carlo simulations. The inversion
profile was twice smoothed.

\section{Discussion and conclusions}
We analyzed high-resolution full-disk velocity data from SOHO/MDI
along the center meridian covering three Carrington rotations from
April to June 1997. Using a Fourier transform and measuring the
frequency shift of the p-mode ridges in the power spectrum, we are
able to obtain a meridional flow profile as a function of
radius/depth. Figure~\ref{fig3} shows that in the Northern as well as
the Southern hemisphere the flow is poleward close to the solar
surface and turns equatorward at around 40\,Mm. Such a return flow was
previously tentatively observed by \citet{BraunFan1998}, analyzing
data from the same period (but only for 8 days instead of 3
months). The flow may again turn poleward in deeper regions indicating
a counter cell, although the data is also consistent with zero flow at
lower regions. Such counter cells are consistent with recent numerical
simulations \citep{Miesch2007}. The near-surface flow being somewhat
larger in the Southern than in the Northern hemisphere may be because of
the inclination of the solar rotation axis \citep{BeckGiles2005}.

\acknowledgements
We thank SOHO and in particular the SOI Team at Stanford University
for making SOHO/MDI data readily available. We are also grateful to the
UK Science and Technology Research Council (STFC) for funding
this research. 

\bibliography{references}




\end{document}